%% file: 0.main.tex
\documentclass[acmsmall,screen]{acmart}

\AtBeginDocument{%
  \providecommand\BibTeX{{%
    \normalfont B\kern-0.5em{\scshape i\kern-0.25em b}\kern-0.8em\TeX}}}

\usepackage{multirow,makecell}
\usepackage{tabu}
\usepackage[per-mode=fraction]{siunitx}
\usepackage{gensymb}
\usepackage{subcaption} 
\usepackage{todonotes}
\usepackage{caption}
\usepackage{xcolor}

\newcommand{\mrv}{}
\newcommand{\rv}{}

\graphicspath{{figures/}{pictures/}{images/}{./}}

\begin{document}

\title{Towards Immersive Collaborative Sensemaking}

\author{Ying Yang}
\email{ying.yang@monash.edu}
\affiliation{%
  \institution{Monash University}
  \streetaddress{Wellington Road, Clayton}
  \city{Melbourne}
  \state{Victoria}
  \country{Australia}
  \postcode{3800}
}

\author{Tim Dwyer}
\affiliation{%
  \institution{Monash University}
  \city{Melbourne}
  \country{Australia}}
\email{tim.dwyer@monash.edu}

\author{Michael Wybrow}
\affiliation{%
  \institution{Monash University}
  \city{Melbourne}
  \country{Australia}
}
\email{michael.wybrow@monash.edu}

\author{Benjamin Lee}
\affiliation{%
  \institution{Monash University}
  \city{Melbourne}
  \country{Australia}
}
\email{benjamin.lee1@monash.edu}

\author{Maxime Cordeil}
\affiliation{%
  \institution{The University of Queensland}
  \streetaddress{St Lucia}
  \city{Brisbane}
  \state{Queensland}
  \country{Australia}
  \postcode{4072}
}
\email{m.cordeil@uq.edu.au}

\author{Mark Billinghurst}
\affiliation{%
  \institution{University of South Australia}
  \streetaddress{Mawson Lakes Blvd}
  \city{Adelaide}
  \state{South Australia}
  \country{Australia}
  \postcode{5095}
}
\email{mark.billinghurst@unisa.edu.au}

\author{Bruce H. Thomas}
\affiliation{%
  \institution{University of South Australia}
  \streetaddress{Mawson Lakes Blvd}
  \city{Adelaide}
  \state{South Australia}
  \country{Australia}
  \postcode{5095}
}
\email{bruce.thomas@unisa.edu.au}

\renewcommand{\shortauthors}{Yang, et al.}

\begin{abstract}
  When collaborating face-to-face, people commonly use the surfaces and spaces around them to perform sensemaking tasks, such as spatially organising documents, notes or images. However, when people collaborate remotely using desktop interfaces they no longer feel like they are sharing the same space. This limitation may be overcome through collaboration in immersive environments, which simulate the physical in-person experience. In this paper, we report on a between-groups study comparing collaborations on image organisation tasks, in an immersive Virtual Reality (VR) environment to more conventional desktop conferencing. Collecting data from 40 subjects in groups of four, we measured task performance, user behaviours, collaboration engagement and awareness. Overall, the VR and desktop interface resulted in similar speed, accuracy and social presence rating, but we observed more conversations and interaction with objects, and more equal contributions to the interaction from participants within groups in VR. We also identified differences in coordination and collaborative awareness behaviours between VR and desktop platforms. We report on a set of systematic measures for assessing VR collaborative experience and a new analysis tool that we have developed to capture user behaviours in collaborative setting. Finally, we provide design considerations and directions for future work.
\end{abstract}

\ccsdesc[500]{Human-centered computing~User studies}
\ccsdesc[500]{Human-centered computing~Interaction devices}
\ccsdesc[500]{Human-centered computing~Collaborative and social computing systems and tools}
\ccsdesc[500]{Human-centered computing~Empirical studies in collaborative and social computing}

\keywords{Collaborative Sensemaking, Virtual Reality}

\setcopyright{acmlicensed}
\acmPrice{15.00}
\acmDOI{10.1145/3567741}
\acmYear{2022}
\copyrightyear{2022}
\acmSubmissionID{iss22main-id9657-p}
\acmJournal{PACMHCI}
\acmVolume{6}
\acmNumber{ISS}
\acmArticle{588}
\acmMonth{12}


\maketitle

\input{1.introduction}

\input{2.related-work}

\input{3.user-study}

\input{4.study-results}

\input{5.conclusion}

\begin{acks}
This research was supported under the Australian Research
Council’s Discovery Projects funding scheme (DP180100755).
\end{acks}

\appendix
\input{6.appendix}

\bibliographystyle{ACM-Reference-Format}
\bibliography{0.main-iss2022.bib}

\end{document}

%% file: 1.introduction.tex
\section{Introduction}

\rv{Remote collaboration continues to grow in popularity, particularly since the COVID-19 pandemic.  Using a wide range of technologies people are engaging in a greater variety of remote collaboration tasks than ever before.
A key example of a task that needs better support for remote collaboration is \textit{collaborative sensemaking}, i.e.,\ working together to make sense of information \cite{paul2010understanding}.}
Standard practice for collaborative sensemaking is video conferencing (e.g., Zoom, \url{https://zoom.us/}) and online tools like shared whiteboards (e.g., Miro, \url{https://miro.com/}), but these do not provide a strong sense of spatial awareness or embodied presence.
Previous research has claimed that Virtual Reality (VR) can overcome these limitations~\cite{witmer1998measuring}. The emerging field of Collaborative Immersive Analytics (CIA) \cite{billinghurst2018collaborative} explores how VR can be used for analysing data in a group setting. 
\rv{For example, \citet{leigh1999computer} describe early work in using VR for collaborative visualisation, and \citet{donalek2014bigdata} provide a summary of collaborative data visualisation using VR. \citet{marai2016interdisciplinary} reviewed several Immersive Analytics projects by groups using a large cylindrical projection environment where people reported on getting more work done in two days than in six months of desktop collaboration tools.}

VR potentially offers an embodied experience closer to in-person collaboration than is afforded by remote collaboration using a desktop interface. However, although CIA systems are promising, there has been little research that compares collaborative sensemaking in VR to a desktop interface and---to the best of our knowledge---none with groups of more than three people at a time. 
\rv{Studying collaboration among groups larger than three is important because it enables different dynamics than with smaller groups, e.g., a group of four may subdivide into two teams of two in order to discuss separately and work in parallel.}

Our research explores whether shared immersive environments can better support group sensemaking tasks compared to a desktop environment.
\rv{
Our study asked groups of four people to arrange and cluster images using remote collaboration via either a VR or desktop interface.
Compared to a desktop condition, we found that teams in VR had similar task effectiveness but that VR provided several benefits, with the teams on each platform engaging in significantly different behaviours in terms of interaction, conversation, coordination and awareness.
From our observation of these different behaviours we derive suggestions for the design of future collaborative VR applications in various aspects, such as communication, notification, navigation, environment and virtual elements, and provide directions for future research. 
}

\rv{
Our study focused on spatial sensemaking tasks, which are fundamental sensemaking behaviour that uses space to make sense of information and manage data, and is a way to externalise thoughts and reduce cognitive affordances~\cite{robertson1998data,andrews2010space}.
Our results showed that, when collaborating with others in VR, people employ various space-use strategies rather than preferring to use the space around them when performed tasks individually, as found in previous research~\cite{batch2019there,satriadi2020maps,lisle2021sensemaking}. 
Interestingly, we still identified common patterns from these different space usages.
These findings can inform future researchers of how the space could be used in collaborative sensemaking, especially when designing and providing automatic spacial layout for organising information and data.
}

We also present a new tool to capture and analyse group behaviour in a collaborative sensemaking task. Past remote collaboration research has found that behavioural measures are often a better measure of technology impact than performance measures~\cite{mccarthy1994measuring,billinghurst2003communication}. 
However, behaviour of groups of more than two people can be complex and difficult to analyse.  We have created new techniques for analysing multiway communication and interactions between group members and objects in the environment.



In summary, our paper makes the following novel contributions:
(1) the first comparative study of a four-person collaborative sensemaking task between a VR condition and a desktop interface condition;
(2) results suggesting benefits to collaborative sensemaking in immersive environments over desktop environments; 
(3) novel systematic measures for understanding and analysing the behaviours of larger groups (four people or more) in an immerse collaborative sensemaking task; and
(4) design considerations for immersive collaborative sensemaking applications and directions for future works.

In the next section, we survey past work informing our research. In Section \ref{sec:study}, we describe our group sensemaking user study and the measures. In Sections \ref{sec:results}, \ref{sec:observation-results} and \ref{sec:discussion}, we present the study outcomes and discuss the results. Finally, we summarise and reflect on findings, limitations and directions for future work. 

%% file: 2.related-work.tex
\section{Related Work}
\label{sec:relwork}

\subsection{Immersive Spatial Sensemaking}

Commercial immersive collaborative techniques and applications are emerging to support sensemaking, such as Microsoft's Mesh (\url{https://www.microsoft.com/en-us/mesh}) or Meta's Workrooms (\url{https://www.oculus.com/workrooms/}), which allow multiple people to join and perform tasks such as presenting data or brainstorming. While there is momentum in the immersive industry to provide collaboration platforms, it is unclear however, how those tools can support sensemaking.
Spatial sensemaking uses spatial organisation to manage data to support understandability and memorability, which has been researched on various platforms~\cite{barsalou1983ad,robertson1998data,cockburn2002evaluating,andrews2010space,endert2012semantics,wenskovitch2020examination}.
\mrv{Immersive techniques extend traditional desktop interfaces into immersive environments, and enable use of embodied representations and intuitive interaction to improve cognition and facilitate sensemaking~\cite{chandler2015immersive, marriott2018immersive}.}

Recent research has studied how individuals and groups perform spatial sensemaking on data and documents in immersive spaces.
\rv{
For example, \citet{batch2019there} found that for immersive data analytics, data scientists tended to use the space in front of them to explore different visualisations while they used the space around them to present their findings to an audience. Similarly, other research also showed that people tended to layout maps and documents spherically around themselves~\cite{satriadi2020maps,lisle2021sensemaking}.
\citet{liu2020design} found that semi-circular layouts in 3D space improve visual search compared to flat layout. 
However, all of this research only examined single-user scenarios.
}

\mrv{Fewer researchers have explored collaborative sensemaking in immersive analytics. }
For example, Lee et al.~\cite{lee2020shared} evaluated a CIA prototype with three simultaneous users, and found that people still organise visualisations in curved layouts around them, but also placed them in flat arrangements to facilitate group discussion and collaboration. 
Most recently, \rv{~\citet{luo2021investigating,luo2022should}} conducted an empirical study to investigate the impact of physical surroundings on spatial arrangement with paired users in Augmented Reality (AR). They found that some users produced cylindrical layouts around them while others used the walls and physical furniture as separation for different clusters.



The main results regarding the use of space for sensemaking in immersive environments are that people tend to arrange information around them in a semi-circular shape or a flat manner. More work is needed to understand how users collaborate and interact with data items, and how this compares to traditional collaborative 2D desktop systems. 


\subsection{Collaboration Proximity and Territoriality}

The collaborative sensemaking process is affected by two spatial factors: the (physical) proximity of collaborators to one another, and the use of territories to manage resources in the workspace. 
Hawkey et al.~\cite{hawkey2005proximity} found that collaboration is more effective when groups physically stand close to each other when using a large vertical display. In contrast, Prouzeau et al.~\cite{prouzeau2017trade} found that participants using separate desktops performed a collaborative path-finding task faster than when using a shared large vertical display. However, they note that the latter produces consistently higher quality results, due to there being more verbal communication in the shared environment. 

How groups manage resources in the common workspace has been shown to influence collaboration styles~\cite{scott2010theory, isenberg2011co, chung2014visporter}. Most notably are the notions of personal, shared and storage territories~\cite{scott2004territoriality}. 
Bradel et al.~\cite{bradel2013large} found that groups who predominantly used shared territories on a large vertical display communicated more than those who mainly used personal territories. 
The partitioning of the shared workspace into territories is natural and rarely requires verbal communication~\cite{tse2004avoiding}. 
Personal territories are commonly directly in front of users~\cite{tang1991findings, kruger2004roles,liu2016shared}, and the shift between personal and shared territories can easily shift as users physically move around the workspace~\cite{tang2006collaborative,jakobsen2014up}.

Previous research shows that personal and shared territories affect collaboration and communication. It is important to explore how such collaboration proximity and territory pattern differ in immersive 3D and 2D flat shared space.

\subsection{Awareness and Presence in Collaborative Workspaces}

\rv{
It is common to perform sensemaking tasks with flat screens, mouse and keyboard setups. For example, using a combination of collaborative production and communication tools such as Miro or Zoom. 
These provide a shared 2D virtual space for the joint manipulation of virtual objects (e.g., documents and images)~\cite{durlach2000presence}, while use of audio and video enables communication, awareness and social presence~\cite{gutwin2004group,sallnas2005effects, bente2008avatar}.
Workspace awareness~\cite{gutwin1996workspace} is key to recognising others' behaviours in a cooperative environment. While collaborators' activities are not difficult to realise in a co-located situation, it is a challenge to maintain awareness via remote collaboration~\cite{gutwin2004group}. 
Social presence, or the sense of ``being with another''~\cite{biocca2003toward}, is desirable as it can enable collaborators to freely converse and seek help \cite{weinel2011closer}.
}

\rv{
Early research has investigated how to increase awareness and presence by showing others' cursors and viewports on a desktop interface such as Telepointers~\cite{greenberg1996semantic}. 
Hauber et al.~\cite{hauber2005social} compared physical face-to-face meetings, 2D desktop, and non-stereoscopic 3D desktop setups for a ranking task. They found that while the physical meeting was superior to the computer-mediated collaborative tools, the non-stereoscopic 3D desktop improved awareness and presence over the 2D desktop setup.
Later, embodied virtual representations have been employed in distributed tabletop and big display applications~\cite{tang2007videoarms,tuddenham2009territorial}.
Zillner et al.~\cite{zillner20143d} presented 3D whole-body telepresence of remote collaborators on a digital whiteboard, and compared this technique to a 2D representation and co-collocated collaboration. Their findings showed that 3D embodiment representation greatly improved collaboration effectiveness. Similarly, Higuch et al.~\cite{higuchi2015immerseboard} employed 3D-processed immersive telepresence over a digital whiteboard and found that, compared to traditional video conferencing, 3D telepresence provided better co-presence and a more enjoyable environment. Pejsa et al.~\cite{pejsa2016room2room} developed an AR technique that projected a remote collaborators' virtual replication onto the physical environment. They found that this life-sized telepresence provided a stronger sense of being together than a Skype video conferencing.
}
 
\rv{
In stereoscopic immersive environments, collaboration and presence is usually enhanced by the use of embodied avatars~\cite{gerhard2004embodiment,piumsomboon2018mini,ens2019revisiting}.
While realistic full body avatars are favoured for collaboration, using non-realistic avatars still produces a good sense of presence and awareness~\cite{steed1999leadership,heidicker2017influence, yoon2019effect,lee2020shared}.
Notably, Cordeil et al. \cite{cordeil2016immersive}, in a collaborative visualisation task, found no difference in communication and perceived sense of presence in a CAVE environment (where participants see each other physically) compared to a pair of connected VR headsets.
However, direct comparison of collaboration in immersive VR to using a desktop interface remains to be addressed, and research especially needs to be done on whether users collaborate, and perceive awareness and presence differently in these two scenarios. 
}

\subsection{Collaboration Measures}

While previous research has studied and measured how people use online collaborative tools for group work and educational purposes (e.g., \cite{roberts2006evaluation, hernandez2019computer, bulu2012place}), and has focused on modelling collaboration behaviours~\cite{gutwin2000mechanics,baker2002empirical,janssen2007visualization,meier2007rating,wang2017use,chen2018role}, little research has investigated the comparison between non-immersive and immersive platforms for sensemaking.

Some research has investigated collaboration in a hybrid system with one person in VR and two on desktop interfaces~\cite{tromp1998small}, and a hybrid of VR, AR, desktop and wall display with three users~\cite{cavallo2019immersive}, but none of these made a comparison between different interfaces and quantified user behaviours and collaboration. 
Radu and Schneider~\cite{radu2019can} proposed an AR system for physics learning in pairs and compared different levels of virtual involvements, from non-AR to including all AR elements. However, their measures mainly focused on learning outcome, user attitude and oral communications.
Billinghurst et al.~\cite{billinghurst2003communication} conducted series experiments to compare paired users in face-to-face, projection screen and AR settings, and proposed a set of 
measures covering oral and gestural communications, while the quantification of user interactions with virtual contents during collaboration was still missing.
\mrv{
This previous research shows that user experience and behaviours have yet to be systematically measured in immersive environment collaborations, and how different these behaviours could be in VR and with a desktop interface is still unknown.
}

\subsection{Summary}

\rv{
Taken together, 
to the best of our knowledge, there has been no comparative study using non-immersive and immersive environments for sensemaking in groups of four, and systematically measuring the similarity and difference of user interactions, collaborations, communications and space usage in these environments. Thus our research is novel, and addresses an important research gap. In the next section we describe the user study developed to explore this space.
}





%% file: 3.user-study.tex
\section{User Study}
\label{sec:study}

Our study aimed to explore how people use a 3D immersive space for collaborative sensemaking when organising images. We wanted to investigate the advantages and disadvantages of collaborating and communicating in a 3D immersive environment compared with collaborating via a 2D desktop tool and communicating using Zoom.

\subsection{Study Design}

We set up two conditions: (1) a remote VR collaboration system which allows participants to jointly arrange images (\textit{VR}), and (2) a remote desktop collaboration system that allows participants to arrange images together while on Zoom (\textit{Desktop}). We used a between-subjects design for the study where each group of four completed the study with one of the conditions. In the \textit{VR} condition, participants were each in their own room with at least $4\times 4$ metres of space, sound isolated from one another but in the same building.  In the \textit{Desktop} condition (conducted during a strict COVID-19 lockdown), participants joined from home using their own computer.

\subsection{Tasks and Data Set}

To investigate sensemaking for image organisation, we designed tasks requiring participants to arrange a set of images into groups around a set of affective labels, i.e.,~emotion terms such as ``glad'', ``distressed'', etc.  In both \textit{Desktop} and \textit{VR}, images and labels were on moveable ``cards" that could be freely placed within the environment.
Participants were required to discuss their feelings about the images and reach a consensus to place the images near the labels with which they felt the images had the strongest correspondence.

We used the Open Affective Standardized Image Set (OASIS)~\cite{kurdi2017introducing} for the tasks, which contains 900 images. Each image in the data set has two dimensions: \textit{valence} (the degree of pleasantness or unpleasantness) and \textit{arousal} (the degree of excitement or calm), where the images' values in these dimensions have been rated by 822 participants. \rv{These two dimensions of emotion served as the ground truth for our image grouping task to measure performance accuracy.} We designed four tasks and varied the number of image groups and images for each task. The first task, which had the least groups and fewest images, was considered as a trial task to allow participants to get used to collaborating in the shared environment used for their condition. The number of groups in the other tasks ranged from 4 to 7, and the number of images in each group ranged from 3 to 8. We plotted the OASIS image distribution in a valence-arousal matrix to select the images for each task and tried to maximise the distance between the groups and minimise the distance between the images within a group.

The emotion terms used as group labels were selected from the affect model from~\citet{russell1980circumplex}, which organises the terms by valence-arousal. We matched the distribution of the images in the valence-arousal matrix to the emotion word coordinates to select the terms for representing each image group. To avoid a learning effect between tasks, we selected different terms and images for each task.

\begin{figure}
    \includegraphics[width=\textwidth]{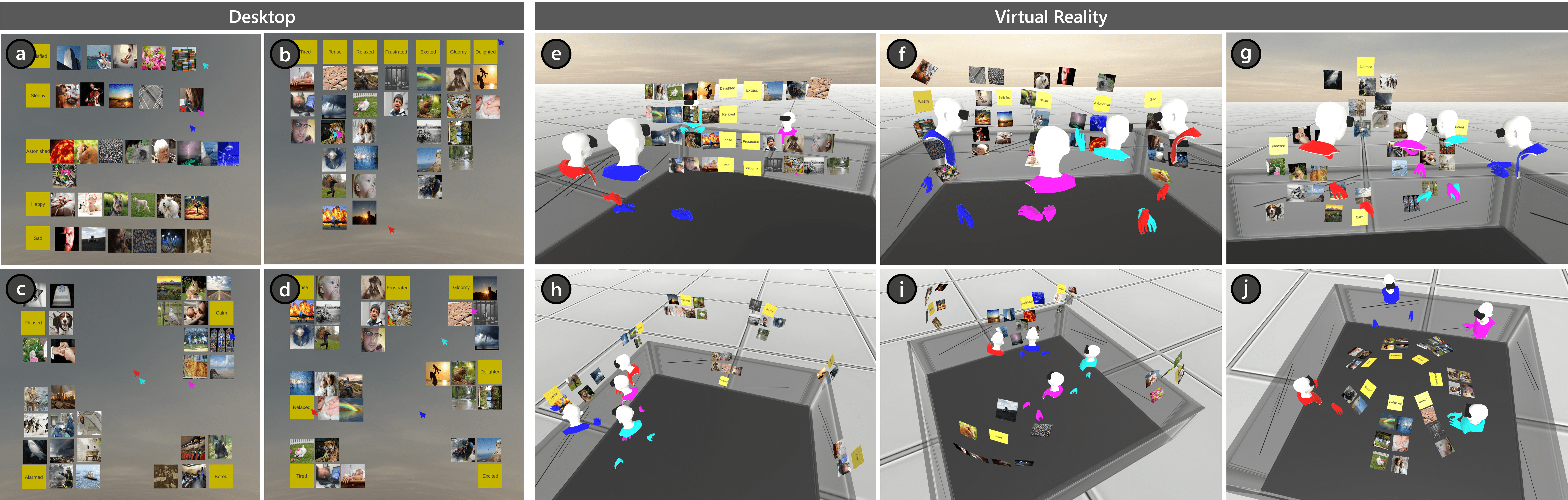}
    \caption{Setup of VR and Desktop conditions. Participants were presented as embodied avatars in VR. On the desktop they see others' mouse cursors in the platform while communicating via conferencing video calls (omitted in the figure). Name tags were visible to participants but are hidden in the figure for anonymity. Individual images show sample layouts created by participants. For Desktop: (a) \textit{Rows}, (b) \textit{Columns}, (c) \textit{Clusters} in corners, and (d) \textit{Clusters} around canvas border. For VR: (e) \textit{Rows} (f) \textit{Columns}, (g) \textit{Clusters} on a 2D ``panel'' (h) \textit{Clusters} on multiple 2D ``panels'', (i) spherical \textit{Clusters}, and (j) \textit{Convex hull}.}
    \Description{figure description}
    \label{fig:teaser}
\end{figure}

\subsection{VR Condition}
\label{sec:vrcondition}
Our \textit{VR} condition allows participants to place the image and label cards hanging freely within a 3D space. 
The VR system was implemented with the Unity3D engine. We used the standalone Oculus Quest 2 VR headsets and controllers in the study.  The headsets are self-contained and require no cord connection to a PC, allowing participants to freely move and rotate their bodies in the space.  The controllers have a trigger that allows participants to naturally grasp and manipulate objects with either hand.
The locations used in the VR condition were empty rooms with a cleared area of $4\times4$\SI{}{\metre} within which each participant had complete freedom of movement.

Initially, the image cards were distributed evenly in the room at waist-height facing up, with randomly generated $x$- and $z$-axis positions and 360\degree{} rotations around their $y$-axis. The same approach was used to distribute the label cards.   Participants were randomly distributed in the space when they started the tasks (by walking to press a ``START'' button displayed at a random location within the space).

To find the most suitable size for images and labels we conducted a pilot study with four participants with varying image and text sizes.  Based on participant feedback, our final study design used image and label cards measuring $20\times20$~\SI{}{\centi\metre}. The label font size was chosen such that the longest label width filled the card.  We used ``LiberationSans SDF'' in the TextMeshPro package as the font style and used the most common yellow colour (R: 255, G: 237, B: 45) as the background for labels (such that they resembled `post-it notes').

The controllers were represented as virtual hands in the environment to give a sense of self-presence to participants. To allow participants to focus on the spatial arrangement, we kept interaction as simple as possible.
Participants could freely move or rotate objects with a simple grip interaction activated with the controller trigger.  When the trigger was pressed, participants would see their virtual hands perform a pinch gesture, and visual feedback of a thick blue outline was provided for objects under manipulation.
Since we were most interested in natural ``embodied'' interaction, we did not support ranged interactions.

Participants could see their collaborators as embodied avatars in the system. We used the Oculus Avatar SDK to implement the avatars, which show head, upper body and hands (as seen in Figure~\ref{fig:teaser}). To distinguish different collaborators, the head and hands were given unique colours, and name tags were displayed above the avatar's head and hands, always facing the viewer's camera.  To support the awareness of collaborators' activities, we provided head gaze by casting a pointing ray from the centre of the avatar's two eyes. Whenever a pointing ray hit an image or label card, a reticle with a name tag was shown at the hit point. 

To achieve remote collaboration and communication in the VR system, we used the Photon Unity Networking framework to synchronise the movement of avatars and objects. The Photon Voice package was used to transmit participants' voice through the network. We implemented 3D spatial sound for the audio of the VR system, so participants could hear the voices of nearby collaborators as loader than the those further away in the shared virtual space. We used a customised logarithmic roll-off curve for the spatial sound, with maximum audible distance of 4 meters.

\subsection{Desktop Condition}

For the Desktop condition, we developed a system with the functionally of the VR system, inspired by the Miro online whiteboard, but with the minimal functionality required for the study tasks. The desktop system was implemented using Unity3D and run on Windows or macOS. The prototype presentation used a windowed mode set to a $2560\times1600$ pixels resolution as the default window size. The application window was resizable. We asked participants to maximise the application to fit their screen. A square 2D canvas was displayed on the centre of the application and used the full window length.

Like in a Miro board, participants in the Desktop condition could see the positions of their collaborators' mouse cursors (see Figure~\ref{fig:teaser}). 
\mrv{Visualising users' mouse cursors in shared space is a conventional approach to support group awareness ~\cite{engelbart1968research}.}
\mrv{This cursor representation is the same as the mouse cursor used in typical desktop systems, familiar for computer users.}
\mrv{Also, it is simple but effective enough to share collaborators' location and movement on the limited desktop screen~\cite{greenberg1996semantic}.}
Each mouse cursor had a unique colour and a name tag above it. The movement of mouse cursors and the image and label cards were synchronised using the Photon Unity Networking framework.

We used the same approach as the VR condition to initially distribute the images and label cards on the canvas. The $x$- and $z$-axis positions of the objects in the VR condition were projected to the $x$- and $y$-axis positions on the desktop canvas. Since people look at the screen from a fixed orientation, the rotation of the objects in this condition is fixed to the orientation of the screen. 
We conducted a pilot study and found that the most suitable scale of the (square) image and label cards as a fraction of canvas width was 0.067. 
To avoid extra navigation on canvas, since the cards can be clearly seen with a normal sized 23 inch monitor, we did not provide zoom in/out functions in the application.
We used the same font style, font size relative to card size and label note background colour as the VR condition.

Participants used a mouse to interact with the desktop application. The only interaction required in the study is \textit{moving} objects. They could move image or label cards by left-clicking on them and dragging with the mouse. As with the VR condition, during interaction the image or label card was highlighted with a blue outline.

\mrv{During the study, participants were connected via Zoom video calls, and their voices were transmitted via Zoom, so they could easily speak to one another.}
\mrv{As we want to compare embodied VR and traditional desktop environment, as a convention, the loudness of users voice was naturally controlled by the distance from a user to the camera and microphone.}

\subsection{Participants}

We recruited 40 participants (13 female, 27 male), resulting in 20 people (5 groups) for each condition. We balanced the age and gender for the VR condition (6 female and 14 male, aged 18-34, M = 26.5, SD = 4.2) and the Desktop condition (7 female and 13 male, aged 18-44, M = 25.8, SD = 5.3). In the VR condition, 20\% of the participants self-reported no experience in VR/AR, 55\% reported having 10 hours or less of experience, and 25\% reported having more than 10 hours experience. Only three participants reported using social VR/AR systems before. The participants in the Desktop condition were asked about previous experience with video conferencing and desktop collaborative tools (e.g., Miro, Google Docs). Over 90\% of participants reported frequent usage of video conferencing, and 55\% reported having used desktop collaborative tools. We offered the participants gift cards for their participation. The experiment lasted 45 minutes.

\subsection{Procedure}

Each condition followed the same general procedure. In the VR condition, to prevent hearing each other, the participants were in separate rooms, each of which had an empty space of at least $4\times4$~\SI{}{\metre}. The participants communicated via the VR system. They were supervised physically by two experimenters and also monitored via Zoom video calls set up in each room. For the Desktop condition, due to COVID-19, the study was run completely online. The participants communicated via Zoom calls, supervised by the experimenter.

The participants needed to enter their names to log into the VR/Desktop system. After login, the participants completed a training protocol individually before meeting their collaborators. The training guided the participants to practise grabbing, moving and placing objects in the system. In the VR condition, the objects were placed at particular positions to force participants to walk around and become familiar with the environment. After training, the participants were shown task instructions. When all four participants finished reading the instructions, they joined the meeting room together---the VR/Desktop environment where they would complete the tasks. In the meeting room, the participants were asked to introduce themselves and do an ice-breaker activity: sharing which animal they wanted to be if they could. When the participants finished the ice-breaker activity and were ready for the tasks, they asked the experimenter to show them the first task. The experimental tasks were fully controlled by the experimenter, who used a server application to control the progress of the study and communicate with the participants. When the participants finished each task, they needed to ask the experimenter to change the task for them. 

In the first task of one group, the participants finished the task without any discussion or collaboration, and so the experimenter prompted them to review the arrangement of the images.  All other groups naturally discussed and collaborated on all tasks.

After the study, all participants were asked to complete a questionnaire regarding demographic information and their previous experience with VR, video conferencing systems and collaboration tools. They were also asked to complete a post-experiment questionnaire, answering short-answer questions about their strategies for image arrangement and collaboration during the study and overall feedback, and rating their perception of Social Presence on a 7-point Likert scale from strongly disagree to strongly agree~\cite{harms2004internal}.

\subsection{Data Collection}

The server application used by the experimenter also collected the study log data: time stamps of actions, positions of image and label cards, and participants' head and hands/mice positions. In addition, we developed an audio recording system to record participants' voice. Four audio recording systems were run during the study, each followed one participant to record their audio.

The Desktop/VR systems that were run by the participants were connected to a Photon server used by the experimenter on a desktop PC, with high-speed Ethernet to minimise the latency of the networking. The server application includes the ability to replay the visuals and synced audio from experimental sessions to facilitate data analysis.

\subsection{Measures and Research Questions}
\label{sec:measures_and_rqs}
Our experimental design is exploratory rather than using formal hypothesis testing.  Therefore, we collected as much data as possible to support a variety of post-hoc analysis.  In the following, we describe the research questions we addressed and the measures that we analyse to do so. The definitions of measure terms refer to Appendix~\ref{app:definitions}. We also make note, where applicable, of some of our prior expectations.

\rv{
Regarding task performance, we measured accuracy and completion time. 
We explored activities in terms of interaction, conversation and coordination, and analysed engagement with respect to awareness and social presence. 
We also investigated spatial usage and territories. 
}

\rv{
In both conditions, the sources of our analyses came from the system logs, playback of log files, audio recordings and post-experiment questionnaires. 
Firstly, we computed the occurrence of a number of objective measures from the playback of log files and audio recordings to identify conversation contents (e.g., planning or greeting), conversation types (task and social), conversation participants (initial participant and responding participants) and conversation targets (label and image).
\mrv{As these measures were objective rather than subjective, one author went through the playback manually to identify behaviour types of interest.}
\mrv{To increase the rigor of these results, the author took another pass and did a double-check.}
Then we went through the playback and identified criteria for each behaviour, and developed a program to count the occurrence of these behaviours from logs. Some example behaviours are: for interaction (grab and regroup), for coordination (monitoring, protection, etc.), and for awareness (no response, shaking, etc.).
We checked the results captured by the program against a sample of the playback to make sure the results matched the behaviours present in the playback.
For the samples of results we checked, the program accurately counted all occurrences of the behaviours.
The definitions and criteria to identify these behaviours are discussed below.
}


\subsubsection{Task}
\textbf{RQ1:} \textit{Is there a difference in sensemaking task accuracy and completion time across VR and Desktop?}
We did not expect that the conditions would affect the task accuracy. We believed the participants in VR would take more time than the Desktop condition to complete the tasks as they needed to walk physically in the environment. 


\subsubsection{Activity}
\textbf{\textit{\rv{Interaction.}}}
\textbf{RQ2:} \textit{Are there differences in frequency, duration and type of people's interactions with objects across VR and Desktop?}
From our logs we can identify precisely which participants interacted with which objects, where they grabbed and moved those objects to, how many times, and how much time they spent moving them. 
We regard the most significant movements as \textit{regrouping}, where those movements change the grouping (the label to which an image is closest).  
We did not expect any major difference in the number of interaction operations, but one might expect differences in how participants interact with objects during communication or perhaps through embodied decision making (e.g., externalising their thought process by different usage of space for objects being discussed or left for later discussion).
\rv{An important measure of effectiveness of collaborative activities is whether all team members are able to contribute equally to the task. We wondered \textbf{RQ3:} \textit{Do people collaborate more equally in VR or Desktop in terms of interaction?} We used the Gini coefficient~\cite{david1968miscellanea,dixon1987bootstrapping} to do the calculation, which ranges from 0 to 1, with 0 meaning all participants of a group contributed to the interaction equally, and 1 indicating they did not contribute equally---typically one of the participants led the whole group during a task.}





\textbf{\textit{\rv{Conversation.}}}
\rv{\textbf{RQ4:} \textit{Do people spend a different amount of time on discussion in VR than Desktop?} 
We analysed the conversation proportion to explore this question, using the procedure from Cordeil et al.~\cite{cordeil2016immersive}. Same as for interaction, we also explored equality of conversation among participants. \textbf{RQ5:} \textit{Do people collaborate more equally in VR or Desktop in terms of discussion?}}

\rv{\textbf{RQ6:}} \textit{Does conversation differ in VR from Desktop?}
We categorised the conversation into task-related and social conversations.
We expected roughly equal task-related conversation between the two conditions, but for the participants in the VR condition to have more social conversations due to being more aware of the people around them.

\textbf{\textit{\rv{Coordination.}}}
\rv{\textbf{RQ7:}} \textit{What are the differences in coordination between participants across Desktop and VR in terms of communication and action activities?}
From previous studies~\cite{gutwin2000mechanics,janssen2007visualization} and our observations, we identified six behaviours of coordination from the playback and audio recordings: \textit{Planning}, \textit{Assistance (question and exchange)}, \textit{Monitoring}, \textit{Protection}, \textit{Respect} and \textit{Accommodation}.
\rv{For detailed definitions of these terms refer to Appendix~\ref{app:definitions}.}

\subsubsection{Engagement}
\textbf{\textit{\rv{Awareness.}}}
As mentioned, we measured the study audio and visual playback to get a sense of the participants' collaborative awareness to address:
\rv{\textbf{RQ8:}} \textit{Is there a difference in the type and amount of communication between users based on the distance between them?}
To investigate how participants acknowledged where the others were, we measured the relationship between the discussion and the distance between participants. We retrieved three behaviours related to awareness from the visual playback and audio recordings: \textit{No response}, \textit{Think aloud} and \textit{Conversation conflict}. We expected that in the VR condition the participants would communicate more with those close to them in VR, and that there would be no such correlation between mouse cursor separation and communication in the Desktop condition.

Another analysis related to awareness concerns gestures to get others' attention. 
\rv{\textbf{RQ9:}} \textit{Is there a difference in the amount of gestures between the two conditions of VR and Desktop?}  
The program retrieved gestures from the visual playback and audio recordings for the analysis. We expected there would be more gestures in the VR condition than the Desktop condition as the VR condition allowed more embodied presence of others in the environment.

\textbf{\textit{\rv{Social Presence Rating.}}}
\rv{\textbf{RQ10:}} \textit{Do people experience Social Presence differently for VR and Desktop?} 
We gathered the subjective rating of Social Presence to evaluate participants' perception of how aware they were of others during collaboration, using the Harms and Biocca's questionnaire~\cite{harms2004internal}. The questionnaire measures the factors of Social Presence in four sub-categories: Co-presence, Attentional Allocation, Perceived Understanding (Message \& Affective) and Perceived Interdependence (Emotional \& Behavioral), covering the aspects of communication, coordination and awareness. 
We expected that the VR condition would be more similar to face-to-face circumstances and the participants would be less aware of each other during tasks in the Desktop condition.

\subsubsection{Spatial Usage}
One interesting question investigates the spatial usage based on observations. \rv{\textbf{RQ11:}} \textit{Do people use the space in VR and Desktop differently? Do they have different strategies to use these space?}
We analysed the visual replay and plotted participant and object positions over time to address this question.  

%% file: 4.study-results.tex
\section{Behavioural Measure Results}
\label{sec:results}

\begin{figure}
  \centering
  \caption{Table of study results.}
  \includegraphics[width=\linewidth]{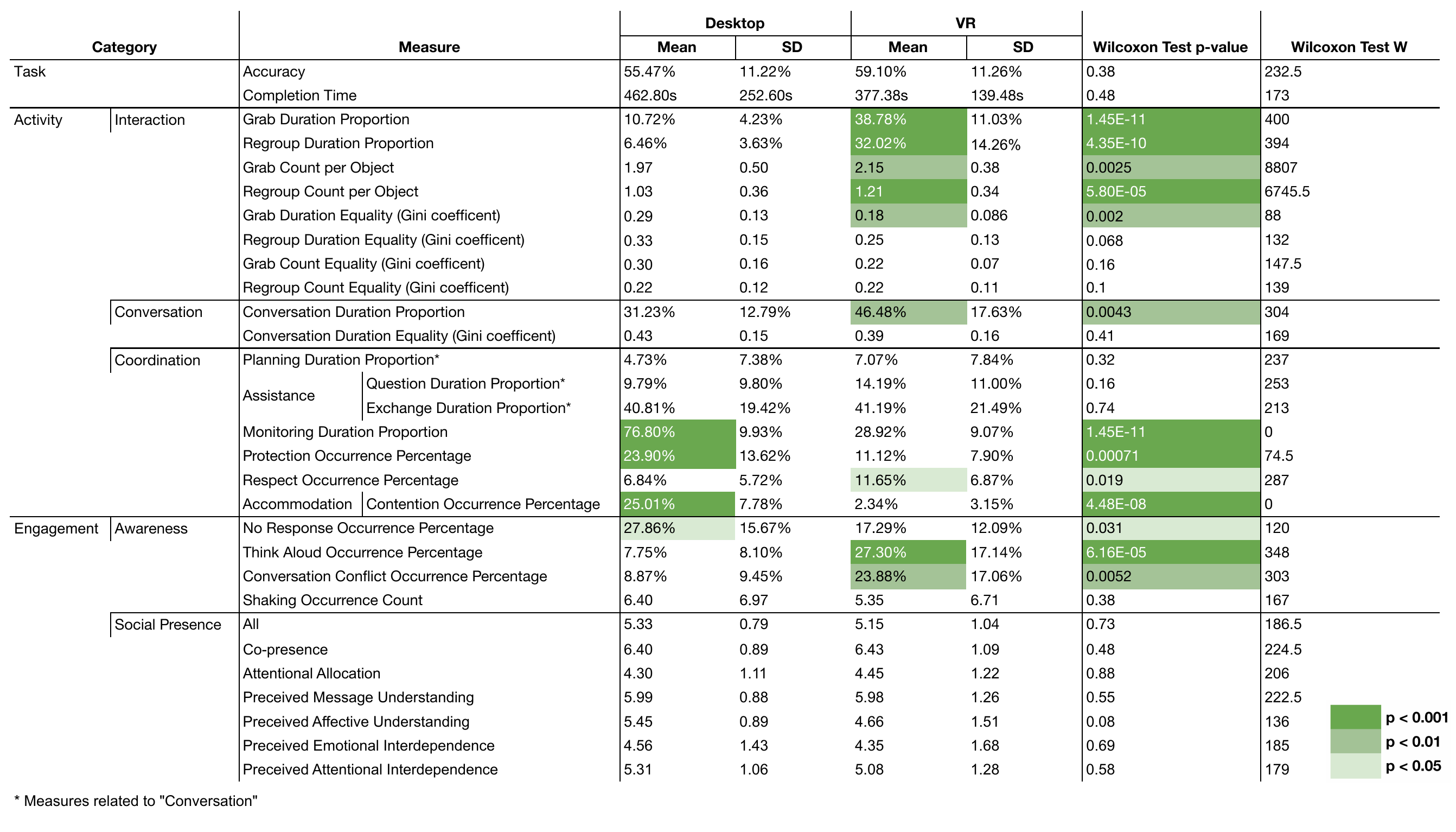}
  \label{table:results}
  \vspace{-10mm}
\end{figure}

The following is a detailed analysis of the various measures in terms of the first nine research questions (RQs). We use the assumption-free non-parametric Wilcoxon's rank-sum test for this analysis~\cite{field2012discovering}. All the statistical results of the measures are presented in Table~\ref{table:results}. A concise summary of the significant results and their relationships to the research questions is provided in Section~\ref{sec:discussion}. Some figures are combined regardless of their order to minimise space.

\subsection{Task}

We did not find significant differences on task accuracy or completion time between two conditions. Regarding \textbf{RQ1} we cannot accept the VR or Desktop condition as being faster or more accurate. However Figure~\ref{fig:task-performance-social-presence} (left, middle) shows that there is a weak trend that the VR condition was slightly faster and more accurate across most tasks.

\subsection{Activity}

\begin{figure}
    \centering
    \includegraphics[height=3.2cm]{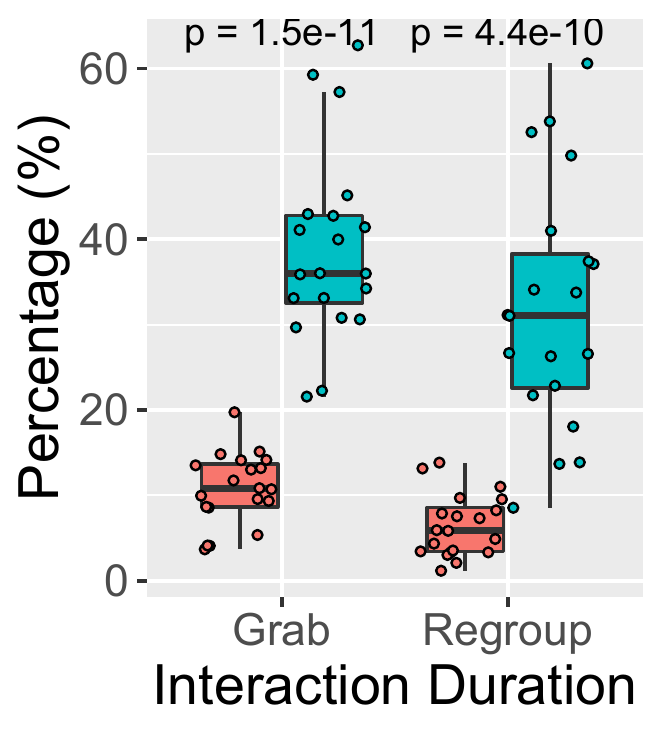}
    \includegraphics[height=3.2cm]{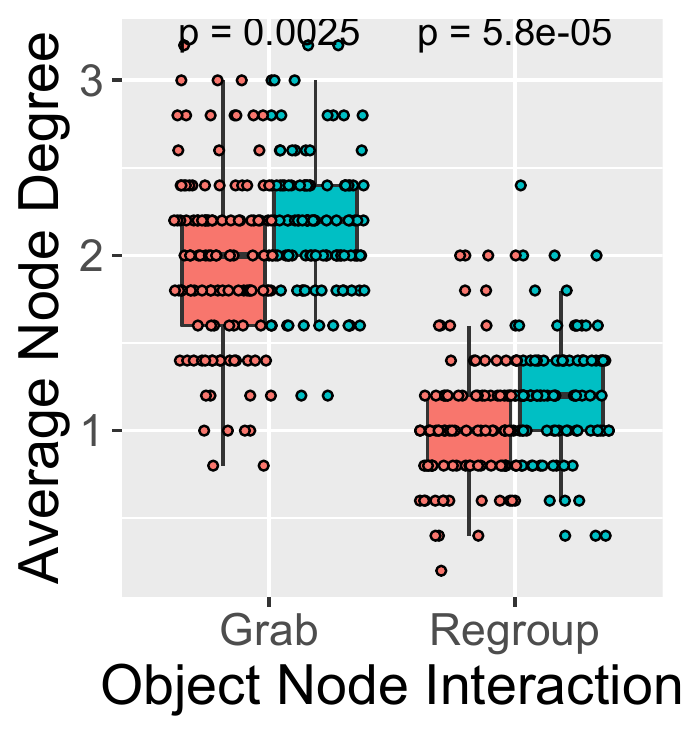}
    \includegraphics[height=3.2cm]{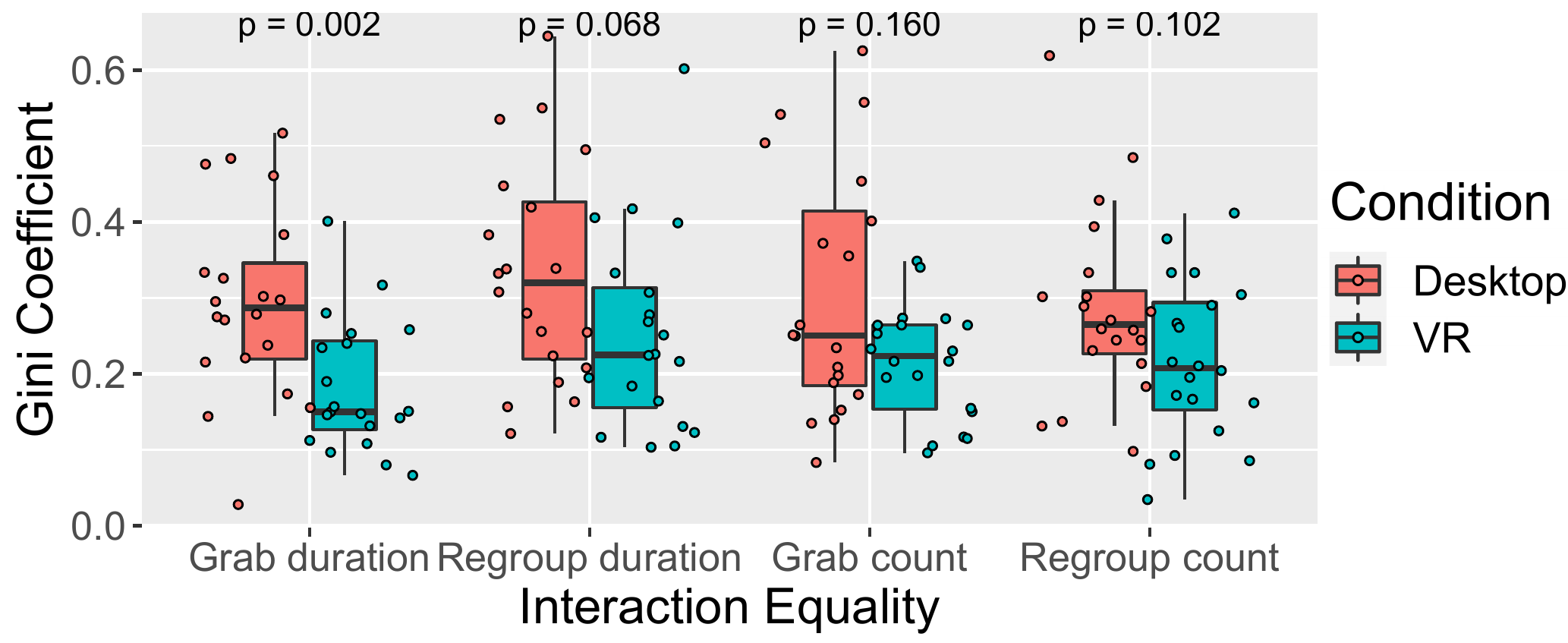}
    \caption{Left: Proportion of time that participants grab and regroup objects (data points represent individual participants). 
    Middle:  The average number of participants grabbing each object, and the average number of participants changing the grouping of each image. The data points represent objects/images. 
    Right: Equality of interactions across tasks (each data point represents one group in one task). 
    \label{fig:interactions}}
\end{figure}

\noindent\textbf{\textit{\rv{Interaction.}}}
Figure~\ref{fig:interactions} (left) shows the proportion of time that participants spent on grabbing (and moving) objects and performing regrouping actions in both conditions (\textbf{RQ2}). 
Participants spent a significantly greater percentage of their time on both grabbing and regrouping in VR than in the Desktop condition.

\begin{figure}
  \centering
    \includegraphics[width=\linewidth]{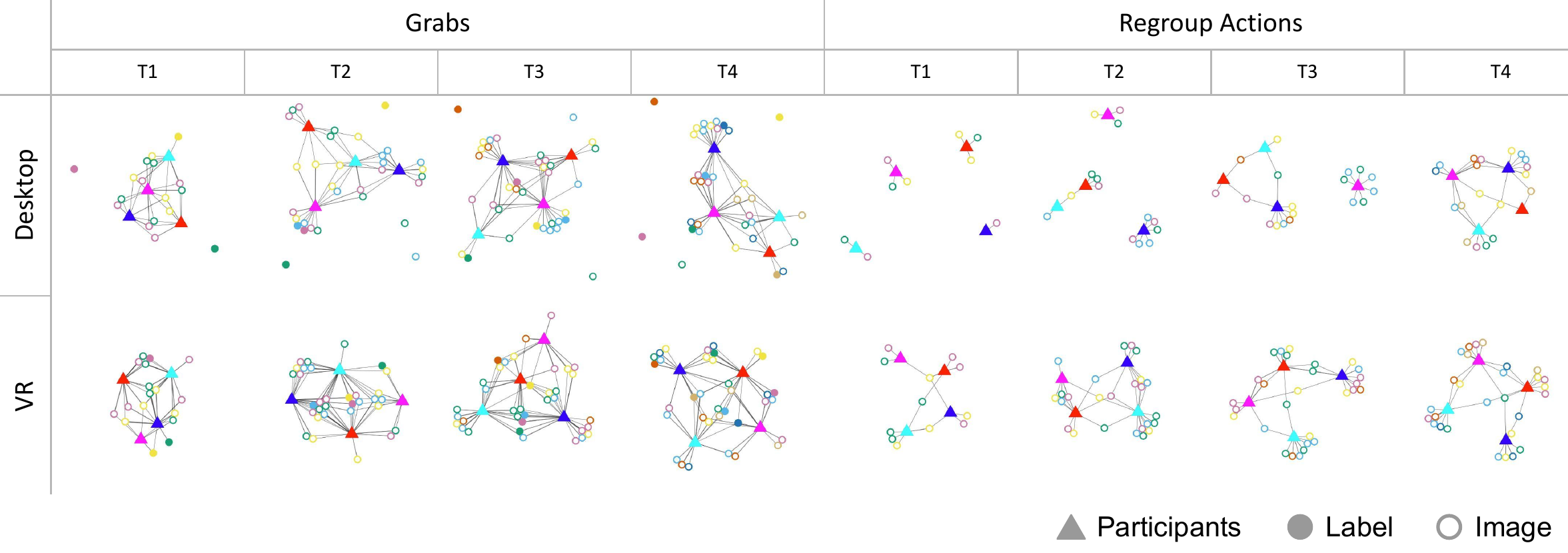}
  \caption{Indicative graphs of participants' interactions with objects in one VR and one Desktop group.  In the ``Grabs" graphs there is a link between a participant 
  and an object (label/image)
  if the object was grabbed and moved by that user.  In the ``Regroup Actions" graphs a link shows the image was moved to a different group by the user. The link thickness indicates the number of interactions. In VR, we see objects are both manipulated and regrouped by more participants than Desktop.}
  \label{fig:grab-graphs}
\end{figure}

To compare participants' interactions with objects across groups and conditions, we computed networks showing which participants grabbed objects (labels and images) and which regrouped images. A sample of the graphs computed for each of the four tasks across one VR group and one Desktop group is shown in Figure~\ref{fig:grab-graphs}.  We saw a common pattern across all groups that in VR participants interacted with more objects than in Desktop and all four members of each VR group tended to interact more uniformly.  This is evident from the more highly connected and symmetric graphs for the rows of VR graphs.  An analysis of the degrees of object and image nodes (Figure~\ref{fig:interactions} middle) also shows the greater degree of interaction in VR.
In VR, objects are interacted with by significantly more participants than in Desktop for both grabbing and regrouping.  

Figure~\ref{fig:interactions} (right) shows the equality of participants' interactions with the objects in each task for both conditions in terms of duration (time spent grabbing and regrouping), count of grabs and regroup operations (\rv{\textbf{RQ3}}).  The only statistically significant difference is in grab duration.
The lower Gini coefficient in the VR condition shows participants in VR interacted with the objects more equally in terms of duration, and significantly more equally than those in the Desktop condition.

\begin{figure}
  \centering
    \centering
    \includegraphics[height=3.2cm]{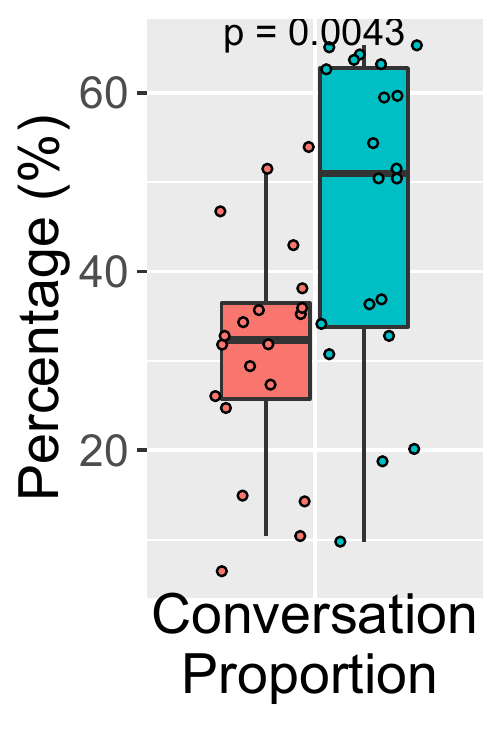}
    \centering
    \includegraphics[height=3.2cm]{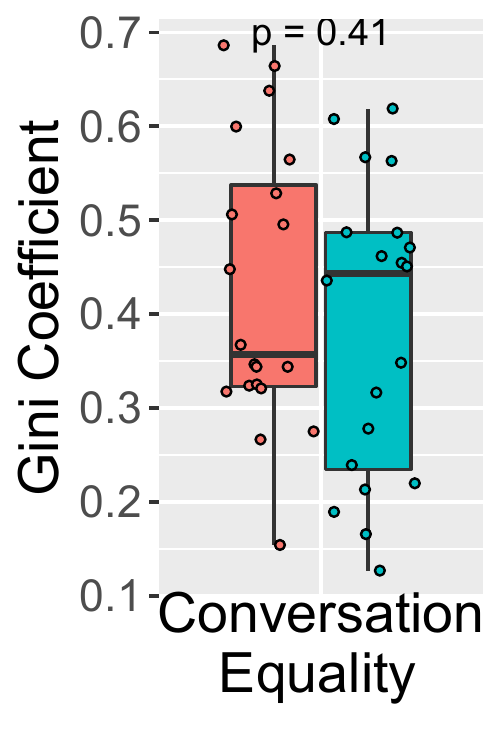}
    \centering
    \includegraphics[height=3.2cm]{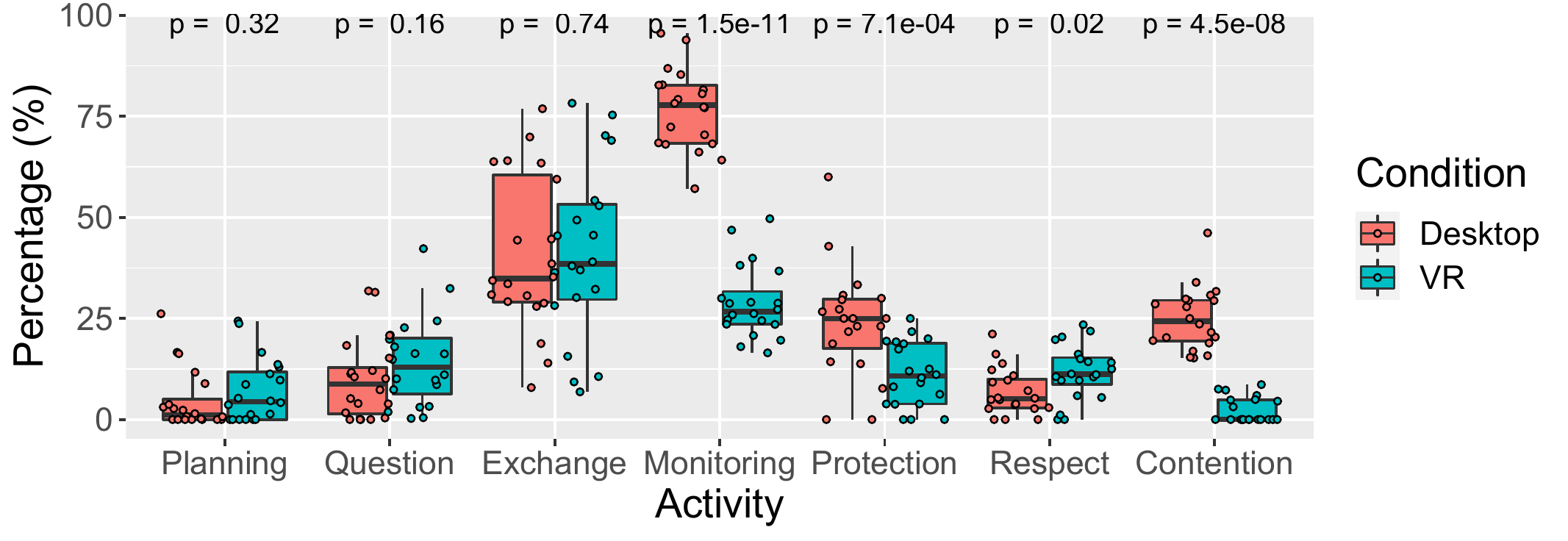}
    \caption{Left, Middle: Oral collaboration performance. Each data point represents one group in one task.
    Right: Coordination activities boxplots. For \textit{Plan}, \textit{Question}, and \textit{Exchange} activities, each data point represents one group in one task, and the percentage is duration of each activity against task completion time. For \textit{Monitoring}, \textit{Protection}, \textit{Respect} and \textit{Contention}, data points represent individual participants. The \textit{Monitoring} percentage relates to duration, while for \textit{Protection}, \textit{Respect}, and \textit{Contention} percentage indicates the number of times that these activities happened to a participant relative to the total conversation/interaction times across all tasks.
    \label{fig:conversation-and-coordination}}
    \vspace{-4mm}
\end{figure}

\noindent\textbf{\textit{\rv{Conversation.}}}
Figure~\ref{fig:conversation-and-coordination} (left) shows the proportion of time that participants have conversations (someone was speaking) in both conditions (\rv{\textbf{RQ4}}). Participants spent significantly more time speaking in VR than in Desktop.
Figure~\ref{fig:conversation-and-coordination} (middle) shows the equality of contributions by participants to the conversation as a proportion of task time in both conditions (\rv{\textbf{RQ5}}). There is no significant difference between the two conditions.
Where there is conversation, Gini coefficient scores between 0.3--0.6 indicate that there is an imbalance between participants in both conditions, although the condition does not effect this significantly.

We identified various topics of social-related conversations in VR: Greeting (2 times), Emotion (4 times), Fun (5 times), Interaction (2 times), Distance (6 times) and Height (2 times) across all groups. The social conversations rarely happened in Desktop, once for Greeting and once for Fun (\rv{\textbf{RQ6}}). Some examples for these topics are as follows. For Fun, participants played with the hand animation and explored it with others. For Interaction, participants tried to ``steal'' objects from others' hands. For Distance, as the participants were remotely located, some participants were surprised that they could pass through others and discussed the feeling of this. Some participants also expressed that the height of some objects placed by other participants were out of their reach. 
The duration of these conversations were mostly between 10 to 45 seconds. 
We identified three topics for the task-related conversations, as shown in Figure~\ref{fig:conversation-and-coordination} (right): \textit{Planning}, \textit{Question} and \textit{Exchange}, which also related to \textit{Coordination} (discussed below). 



\noindent\textbf{\textit{\rv{Coordination.}}}
Figure~\ref{fig:conversation-and-coordination} (right) shows there are significant differences of \textit{Monitoring}, \textit{Protection}, \textit{Respect} and \textit{Contention} (\rv{\textbf{RQ7}}). The participants in Desktop spent much more time on monitoring than in VR. They also tended to protect their work more than in VR. Participants respected others' work more in VR than in Desktop. For \textit{Accommodation}, we found participants exhibited the \textit{Voting} activity once in VR and twice in Desktop, and they tried to interact with the same objects significantly more in Desktop than in VR. 
There is no significant difference of \textit{Planning} and \textit{Assistance} (\textit{Question} and \textit{Exchange}) activities between two conditions (\rv{\textbf{RQ7}}).

\subsection{Engagement}

\begin{figure}
\centering
\includegraphics[width=\linewidth]{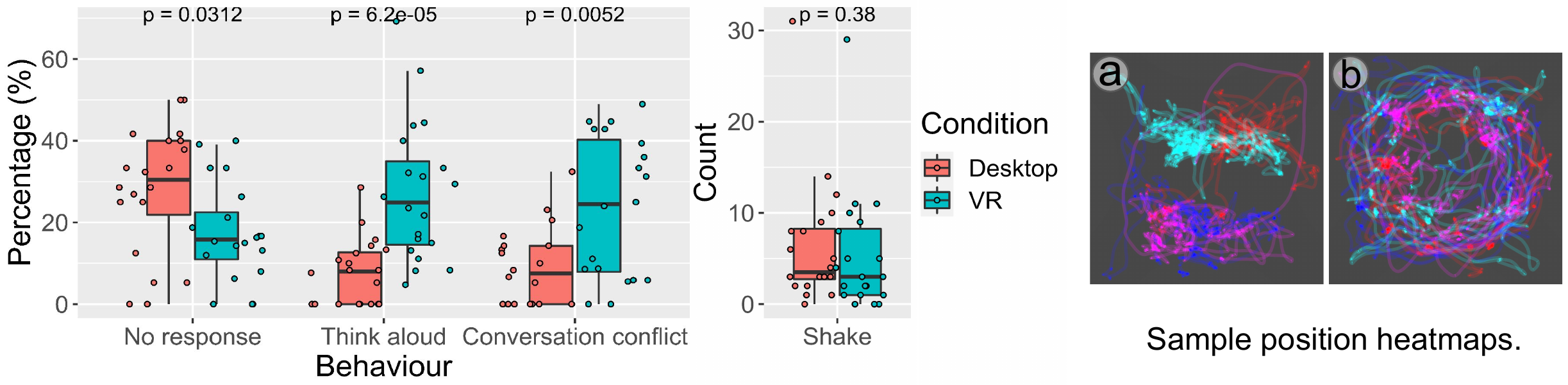}
  \caption{Left, Middle: Awareness behaviours boxplots. 
  The data points for \textit{No response}, \textit{Think aloud} and \textit{Conversation conflict} behaviours represent tasks. The percentage values indicate the relative number of times these behaviours occurred.
  For the \textit{Shake} behaviour each data point represents individuals, and the count value indicates the number of times that participants shook their controller/mouse in all tasks.
  Right: Sample heatmaps of participants' $x$- and $z$-axis positions over time. Each heatmap is generated for one group in one task, and each colour represents a participant.}
  \label{fig:awareness}
\end{figure}

\noindent\textbf{\textit{\rv{Awareness.}}}
\label{sec:coding-of-awareness}
Figure~\ref{fig:awareness} (left) shows there are significant differences of \textit{No response}, \textit{Think aloud} and \textit{Conversation conflict} behaviours. There are more occurrences of \textit{No response} in Desktop than in VR, but less \textit{Think aloud} and \textit{Conversation conflict} happened in Desktop than in VR. 


We produced heatmaps of participants' $x$- and $z$-axis positions to explore the relationship between distance and communication between participants (\rv{\textbf{RQ8}}).
In VR, we observed that the parallel conversations happened wherever the participants were. 
For example, Figure~\ref{fig:awareness} (right (a)) is the heatmap of a group in one task. It shows most of the time the participants worked in pairs (blue and magenta, red and cyan) and stood on the opposite sides of the ``object wall'' they built in the center of the room (see Figure~\ref{fig:teaser} (e)). Besides talking to the person on the same side, participants also talked to the others across the ``object wall". 
The conversations with no responses happened when the other participants stood far away and had some participants close to them, as shown in Figure~\ref{fig:awareness} (right (b)) that the participants freely walked around the room during the task.
The heatmaps for the Desktop condition did not reveal that there was a relationship between the distance between participants (in terms of mouse cursor position) and the communication pattern. 

Two gestures were identified that participants performed to attract others' attention: \textit{Pointing} and \textit{Shaking}. Since \textit{Pointing} happened in every conversation in Desktop condition (i.e., mouse hovering over objects) as the only awareness indication, we did not include it into the discussion. 
The \textit{Shaking} gesture was counted during speaking. In VR, participants shook their virtual hands or objects, and in Desktop participants quickly moved their mouse cursor around on or with objects. As can be seen from Figure~\ref{fig:awareness} (middle) there is no significant difference across two conditions (\rv{\textbf{RQ9}}).







\begin{figure}
  \centering
  \includegraphics[height=3.35cm,trim=0cm 0cm 3cm 0cm, clip]{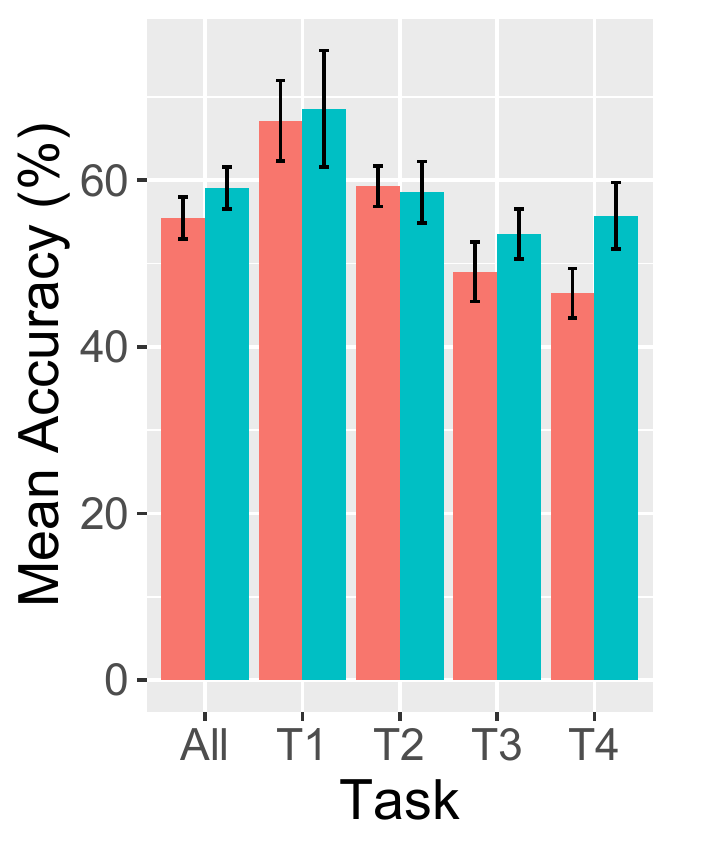}
  \includegraphics[height=3.35cm]{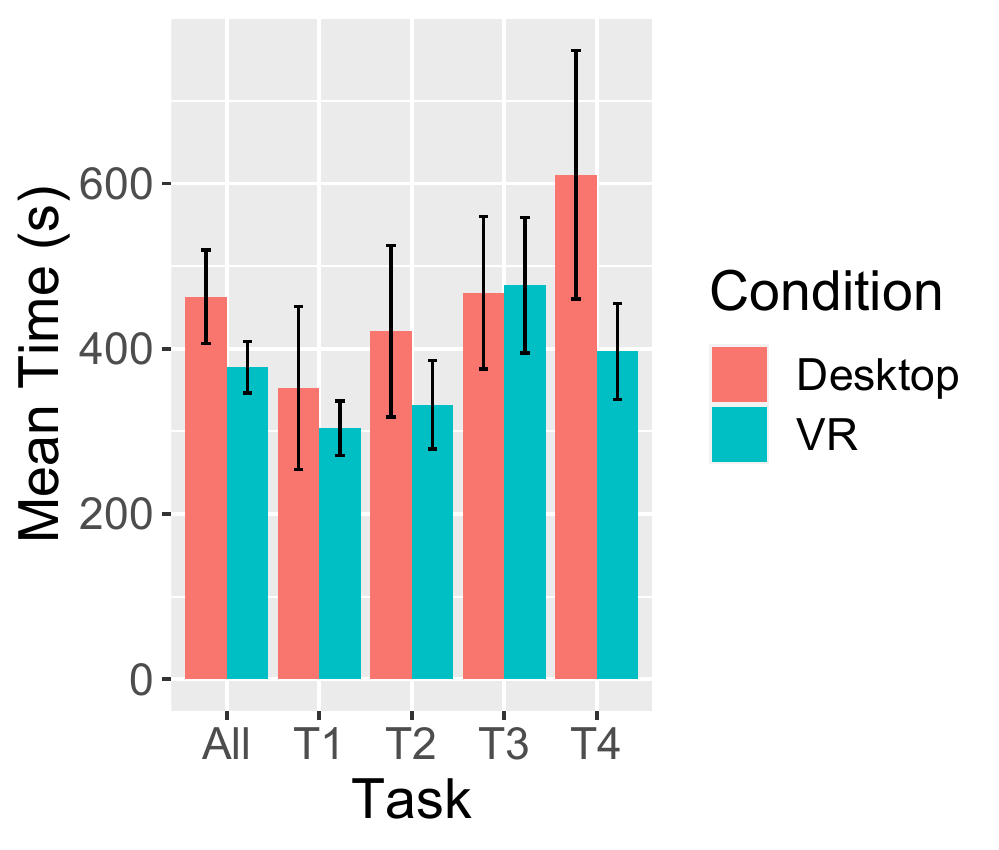}
  \includegraphics[height=3.35cm]{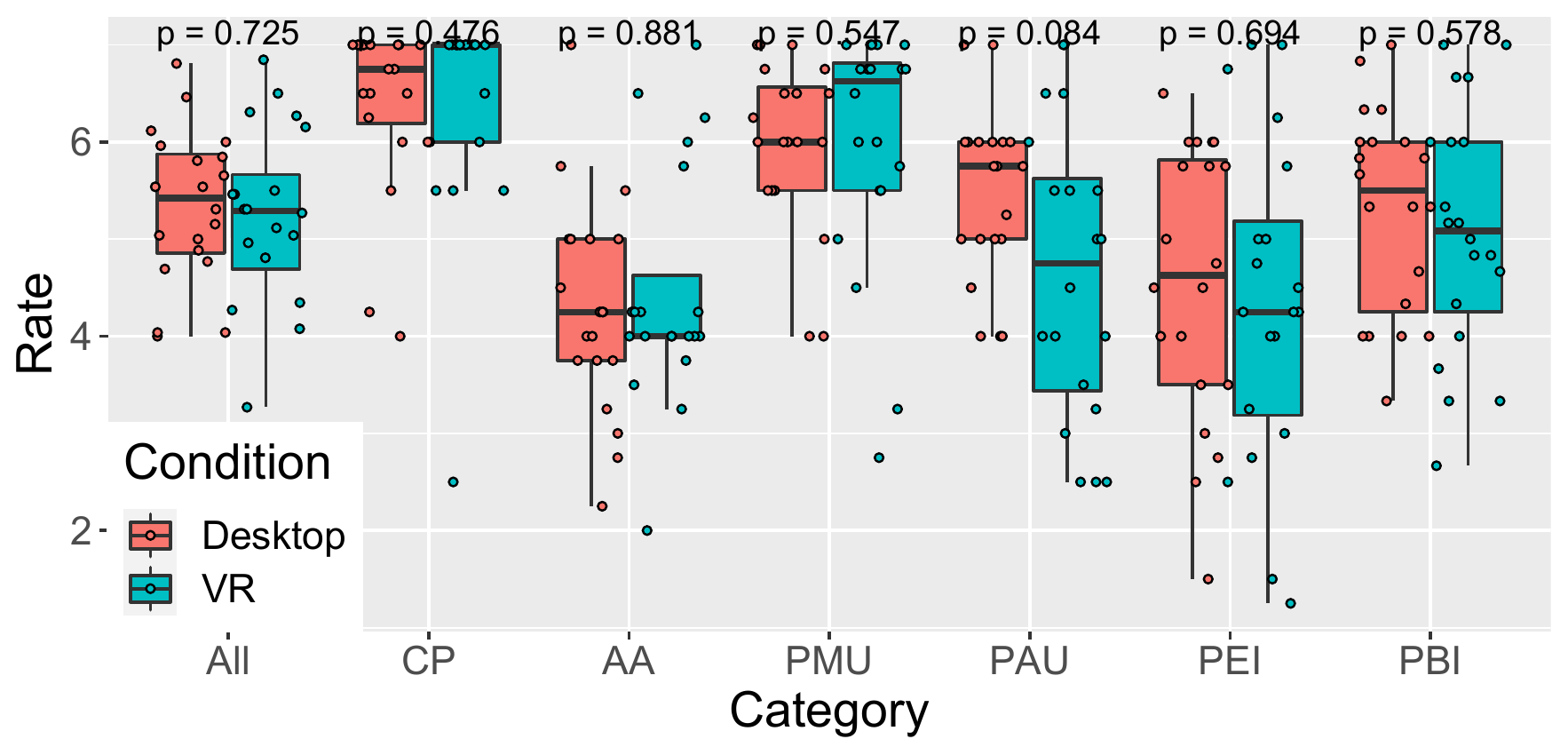}
  \caption{Left, Middle: Task accuracy and completion time across all groups. Right: Boxplots for Social Presence results of overall rating and each sub-categories: Co-presence (CP), Attentional Allocation (AA), Perceived Message Understanding (PMU), Perceived Affective Understanding (PAU), Perceived Emotional Interdependence (PEI) and Perceived Behavioral Interdependence (PBI). The data points represent individual participants. }
  \label{fig:task-performance-social-presence}
  \vspace{-4mm}
\end{figure}

\noindent\textbf{\textit{\rv{Social Presence Rating.}}}
Except for the sub-category Attentional Allocation, the Cronbach's Alpha of other sub-categories ranges from 0.90 to 0.96 in VR and from 0.78 to 0.95 in Desktop, showing high internal consistency. The Cronbach's Alpha of Attentional Allocation is 0.65 and 0.64 for VR and Desktop, which are in an accepted range. Figure~\ref{fig:task-performance-social-presence} (right) shows there is no significant difference in social presence in each sub-category across both conditions (\rv{\textbf{RQ10}}).



\section{Observation Results}
\label{sec:observation-results}

To explore \rv{\textbf{RQ11}}, we analysed the study visual replay in detail in terms of: Strategies, Organisations and Territories,
\mrv{which were widely used to analyse collaboration behaviours in previous research~\cite{bradel2013large,clayphan2016wild,lee2020shared,luo2022should}.}

\noindent\textbf{\textit{Strategies.}}
Participants' strategies for space use were quite similar in both conditions, where the participants divided the space into regions: \textit{grouped region}, \textit{unsure region} and \textit{ungrouped region}. The empty space between the regions tended to be used temporarily by the participants for placing the objects for discussion.

Collaboration strategies were also quite similar across conditions. We identified two main strategies: 
\textit{Strategy~1} Participants worked individually to organise images first (Phase One). Some groups put images they were unsure of in an empty space, then subsequently discussed the images in the unsure pile (Phase Two). When they finished an initial grouping of all images, they started reviewing the groups (Phase Three). This basic strategy was used by four groups in both VR and Desktop. \textit{Strategy~2} Participants gathered all the images to the centre of the space, and discussed and grouped them together. One group in VR and one group in Desktop used this strategy.

\noindent\textbf{\textit{Organisation.}}
\label{sec:organisation}
There were two types of organisation in VR condition: (1) 2D space (two groups) and (2) 3D space (three groups). 
The images in the organisation using 2D space were basically formed into a kind of ``panel'' and these panels were either placed in the centre of the room (see Figures~\ref{fig:teaser}(e) and (f)) or on one side of the room (see Figure~\ref{fig:teaser}(g)). There were three patterns of layout of the objects on these 2D panels, e.g., Figure~\ref{fig:teaser}(e) the images followed the labels and formed \textit{Rows}, Figure~\ref{fig:teaser}(f) the images followed the labels and formed \textit{Columns}, and Figure~\ref{fig:teaser}(g) the images were placed around the labels into \textit{Clusters}. 
We identified three patterns from the organisations using 3D space. The first is depicted in Figure~\ref{fig:teaser}(h), the participants organised the labels around multiple sides of the room and put the images around these labels to form some \textit{Clusters}. Each cluster was organised as a small 2D ``panel''. In the second, the participants fully used the 3D space and organised the labels far away from one another as possible. For example in Figure~\ref{fig:teaser}(i), four labels were put into four corners and one label was placed in the centre of the room. The images placed around them formed a spherical layout in each \textit{Cluster}. For the third, one group created a \textit{Convex hull} and built a kind of ``console'' in the centre of the room (see Figure~\ref{fig:teaser}(j)). The labels were placed around a circle in the centre of the console and the images followed the labels and extended from the centre to the outside of the console.


Spatial organisations in the Desktop condition were more uniform than in VR. We observed three patterns of the layouts in Desktop: (1) \textit{Rows} (one group), (2) \textit{Columns} (two groups), and (3) \textit{Clusters} (two groups). The rows and columns layouts were similar to the 2D ``panel'' layouts in VR condition as shown in Figures~\ref{fig:teaser}(a) and (b). For the cluster layouts, each cluster occupied a separate space on the 2D canvas, either on the corners (when number of labels = 4 (see Figure~\ref{fig:teaser}(c)) or around the border of the canvas (see Figure~\ref{fig:teaser}(d)). 

In both conditions, participants either planned the arrangements (two groups in VR, three groups in Desktop) or instinctively organised the objects without discussion (three groups in VR and two groups in Desktop).  The organisation planning normally happened at the beginning of the tasks, but some groups refined the plan in the middle of the tasks when they found the initial organisation did not work for the task. The groups that had spatial organisation plans usually created more organised layouts in both conditions, such as \textit{Row} and \textit{Column} layouts. When there was no spatial organisation planning, the participants just followed others' strategies. This normally happened in the groups using \textit{Cluster} layouts in both conditions. Interestingly, the VR group that produced the \textit{Convex hull} layout (see Figure~\ref{fig:teaser}(j)) did not plan the organisation. They collected the labels in the center of the room at the beginning of Task 1, and stood in a circle around the labels, facing others to discuss the word meanings. When one participant rotated a label towards themselves to see the text, others used the same behaviour. As a result, they naturally extended the grouping of images outwards from the labels in the center and built a ``console'', and used the same layout throughout the subsequent tasks. 

\begin{figure}
  \centering
  \includegraphics[width=\linewidth]{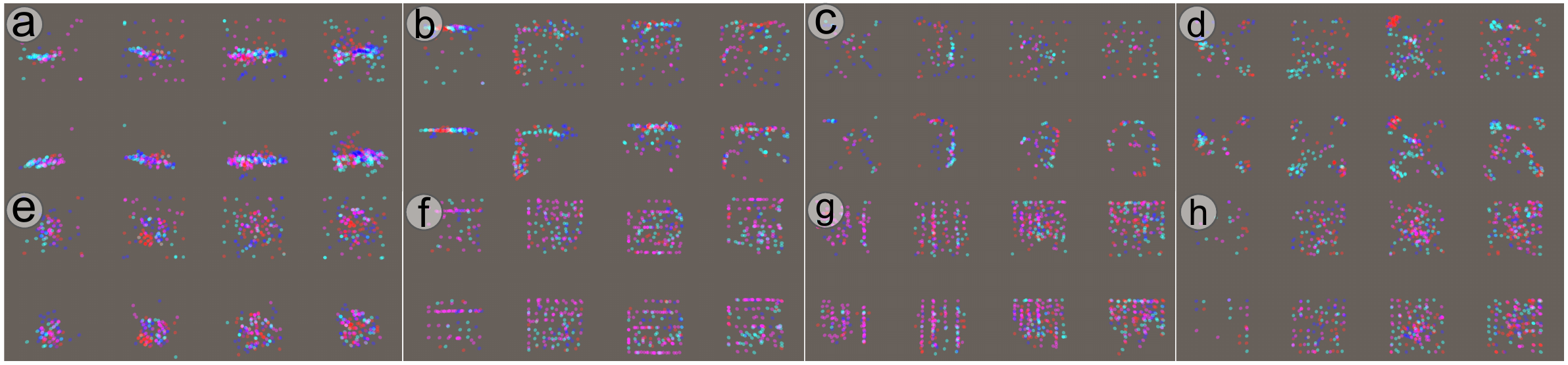}
  \caption{\rv{Sample participant interaction heatmaps that indicate the territories of participants. Two groups of Desktop condition are omitted as their heatmap patterns are similar to (h).} The dots represented the positions when an object was picked/placed by a participant, coloured the same as the participant. Sub-figures (a)--(e) are $x$- and $z$-axis positions of objects in VR and (f)--(h) are $x$- and $y$-axis positions of objects in Desktop.}
  \label{fig:grab-heatmap}
\end{figure}

\noindent\textbf{\textit{Territories.}}
We observed territory patterns from Figure~\ref{fig:grab-heatmap}, which show the positions of objects when picked and placed. Each sub-figure is generated for one group, and each column in the sub-figures represents each task. A dot is drawn when a participant picked (first row in each sub-figure) or placed (second row in each sub-figure) the object. The dots are colour-coded by participants.  Figures~\ref{fig:grab-heatmap}(a)--(e) are the heatmaps for the groups in VR, and Figures~\ref{fig:grab-heatmap}(f)--(h) shows three patterns in Desktop. These territories match the object organisation discussed previously. 

Figure~\ref{fig:grab-heatmap}(a) reveals how participants in VR built a 2D ``panel'' in the centre of the room, and Figure~\ref{fig:grab-heatmap}(b) shows how they built a panel along the sides of the room. Figures~\ref{fig:grab-heatmap}(c)--(e) show use of territories in 3D space. Figures~\ref{fig:grab-heatmap}(c) and (d) depict building a spherical layout, and Figure~\ref{fig:grab-heatmap}(e) is for the ``console'' built in the room centre.

The territory heatmap for Desktop condition is uniform, e.g., Figure~\ref{fig:grab-heatmap}(f) shows the participants built the row layout, Figure~\ref{fig:grab-heatmap}(g) shows is the heatmap for the column layout, and Figure~\ref{fig:grab-heatmap}(h) is the heatmap for the cluster layouts. Interestingly, Figure~\ref{fig:grab-heatmap}(h) column 3 and 4 also reveals that in Task 3 and 4 the participants collected all the objects into the canvas centre at first to make space for organising the grouped images.


\noindent\textbf{\textit{Feedback.}}
The questionnaire prompted participants for their open-ended thoughts and suggestions. For both Desktop and VR conditions, feedback overwhelmingly described the systems as being fun to use and easy to collaborate in. 
For the Desktop condition, 
some comments were made by single participants: auto-alignment tools for the cards; a highlight button to attract attention of group members; and the ability to see each other within the application itself, akin to online services such as Gather (\url{https://www.gather.town/}).
For the VR condition, three participants pointed out that they felt the headsets were uncomfortable after some time. Two participants had requested more communication methods such as symbols or emoji-like responses, particularly to show agreement in group decisions. The remaining comments were given by single participants: poor provenance as there was no way to identify who moved the cards without asking; difficulties caused by height differences between group members; a feature to avoid collisions with each other; and a way to select and move multiple cards at once.

\section{Discussion}
\label{sec:discussion}
\rv{
There are several notable findings from the results of the user study.
}
Firstly, VR prompted more interactions with virtual objects and more communications between participants than Desktop, especially much more social conversations. 
In the Desktop condition, participants spent more time monitoring others' behaviours, but interestingly they had more contention while interacting with objects than in VR. 
VR participants initiated more discussions when they tried to modify others' work, while Desktop participants were involved in more discussions concerning their own work. 
There was more parallel talking in VR than in Desktop, and more no-response conversation happened in Desktop than in VR. 
Detailed discussions on the measures follow. 

\subsection{Task}

The results of accuracy match our expectation discussed in Section~\ref{sec:measures_and_rqs} (\textbf{RQ1}), where the condition did not have effects on accuracy. Unexpectedly, participants in the Desktop condition did not complete the tasks faster.
\rv{
From observation, we found that in VR participants often used two hands to grab two images at the same time, which potentially saved time for moving images. 
Another explanation for the similar completion time is that the space in VR is not large, so participants could easily walk in the virtual room. As the study was designed to examine natural embodiment experience, we did not include extra navigation techniques (e.g., teleportation for VR, panning and zooming for Desktop). It would be interesting to investigate in the future whether the navigation techniques affect task completion time in a large virtual space. 
}

\subsection{Activity}
\label{sec:activity}

\noindent\textbf{\textit{\rv{Interaction.}}}
\rv{
Participants grabbed and regrouped objects more frequently and spent significantly longer time on these in VR than in Desktop (\textbf{RQ2}), and VR participants contributed more equally on grabbing objects (\textbf{RQ3}). 
From observation, the reason for this could be that aligning objects in 3D space requires more effort than in 2D space, so participants adjusted the placement of objects more in VR than on the Desktop to make the organisation neater. This indicates that there is a desire to have auto-alignment mechanisms for objects placed in an immersive environment.
Also, we observed that when there was not enough space to add more images into individual groups, participants spent longer time in VR than Desktop on moving the images one-by-one to make more space for the existing image groups. Providing functions to allow grouping and moving multiple images together will facilitate this object arrangement process.
}
\mrv{Another reason for more interaction and longer interaction time in VR could be that the animated hand representation in VR gave strong self-presence and embodiment sense to participants, which may increase the enjoyment of interactions and stimulate participants to interact more with objects.}

\noindent\textbf{\textit{\rv{Conversation.}}}
\rv{
Participants spent significantly more time speaking in VR than Desktop (\rv{\textbf{RQ4}}). This is surprising but could be attributed to VR increasing the sense of physically being in the space and working with other people.
However, there is no significant difference in conversation equality of participants between the two conditions (\rv{\textbf{RQ5}}).
As expected, we observed that the task-related conversations were similar in the two conditions, and there were much more social conversations in VR than Desktop (\rv{\textbf{RQ6}}). 
This might suggest that people in VR feel more like speaking one-to-one rather than one-to-many in a Zoom like interface. Also, VR might provide a more natural and less professional shared space feeling than video conferencing, as the latter has already been widely adopted in various professional occasions.  One possibility to shape a more professional workplace in VR could be providing professional scenes, avatars and objects to simulate an office or conference venue.
Another interesting finding is about the conversation on height. In VR when there was a great disparity in the heights of the participants, sometimes the short participants commented there were some images outside their accessible range, and then others helped them to reach or place the images. While providing distance interaction can solve this issue, another possibility is employing navigation techniques to ``scroll'' the whole virtual space, which not only supports interaction but also give everyone the power to glimpse the whole space.
}

\noindent\textbf{\textit{\rv{Coordination.}}}
\label{sec:coordination}
Regarding coordination activities (\rv{\textbf{RQ7}}), participants in the Desktop condition monitored others' behaviours and subsequently protected their own works more than in VR.
\rv{It is harder to get an overview of the space in VR than Desktop, especially to notice what is happening outside your own field of view in an immersive environment. One possibility is to provide notifications of changes made by others to inform users.}
Participants in VR respect others' work more than in Desktop. Although there is no difference in the social presence perception rated by participants across the two conditions, we suspect that the greater presence and awareness of working with others made the participants respect others more in VR than in the Desktop condition. Also, the lack of presence led to Desktop participants having more contentions on interaction with objects than in VR.  

\subsection{Engagement}
\label{sec:engagement}
\noindent\textbf{\textit{\rv{Awareness.}}}
As shown in Figure~\ref{fig:awareness}, there were more \textit{Think aloud} and \textit{Conversation conflicts} in VR than Desktop, as separate parallel conversations can occur in the VR environment but are impossible in Zoom calls. 
However, we did not observe any relationship between distance and communication for awareness (\rv{\textbf{RQ8}}).
As discussed in Section~\ref{sec:coding-of-awareness}, the parallel conversations in VR happened regardless of the collaborators' positions in the virtual space.
\rv{This is an interesting phenomenon as the spatial sound did not work as in the physical world where people tend to only talk to others standing close to them. To support a better experience of parallel conversations, VR systems could provide sound isolation mechanisms to allow users to choose whom they want to talk with.}
There were more occurrences of \textit{No response} occurring in Desktop than VR, which could be due to people in Desktop being uncertain who a question was directed towards or perhaps because of lower engagement. 
Also, there were slightly more \textit{Shaking} behaviours in Desktop than VR (\rv{\textbf{RQ9}}), which is contrary to our expectation. We suspect that it was more difficult for participants in Desktop to get others' attention than in VR.

\noindent\textbf{\textit{\rv{Social Presence Rating.}}}
Surprisingly, participants did not give a significantly lower rating on Social Presence in Desktop than VR (\rv{\textbf{RQ10}}) as we expected.  Perhaps people are now so familiar with remote Desktop collaboration that they feel relatively comfortable communicating, despite the impediments we observe above.

\subsection{Observations}
\label{sec:observations}
\noindent\textbf{\textit{\rv{Spatial Organisation.}}}
Regarding \rv{\textbf{RQ11}}, although the spatial organisations in Desktop are neater, we can find common patterns of the layouts across two conditions, like \textit{Rows}, \textit{Columns} and \textit{Clusters}.
\rv{This suggests that even in immersive 3D space, users tend to organise objects as conventionally as they do in 2D space. Future research could be done to design and provide automatic layout options in immersive sensemaking systems to facilitate object organisations.
We also found diverse space usage from different groups in VR, which is different from previous studies~\cite{batch2019there,satriadi2020maps,lisle2021sensemaking,lee2020shared}. This suggests that the nature of tasks, devices, and individual differences need to be considered when designing an automatic organisational layout system.
}
We did not observe which spatial usage and organisations facilitated or impaired task performance and collaboration, and we note that this could be due to the limited number of appearances of each organisation pattern. 

\noindent\textbf{\textit{\rv{Territories.}}}
There is no clear boundary of individual working spaces in either condition. Although there were individual and collaborative working phases, participants freely walked within the room in VR or moved their mice on the canvas on the Desktop to interact with objects or communicate with others during the whole task. 
\rv{This leads to an interesting question for future research: should there be functions to claim individual territories in the shared space and would this positively or negatively impact collaboration?}


\noindent\textbf{\rv{Other Concerns.}}
Although we did not ask for physical workload rate from participants, participants reported discomfort with VR headsets. However, the physical affordance seems not to impede participants' engagement and activity in the VR condition.

\subsection{Design Considerations for Immersive Collaboration}

\rv{
Here we summarise from the detailed analysis above the key considerations that may inform design of and research into future immersive collaboration systems:
}

\noindent\textbf{Communication} (\mrv{based on the discussion on \textit{Awareness} in Section~\ref{sec:engagement}}):
active (manual or automatic) control over participant volume levels is essential to support parallel conversations.  In our \mrv{VR} system we used spatial audio with a custom logarithmic roll-off (see \ref{sec:vrcondition}). Participants indicated they wanted greater control, so either better automatic sound attenuation needs to be developed or systems need to consider providing functions to mute certain collaborators.

\noindent\textbf{Notification} (\mrv{based on the discussion on \textit{Coordination} in Section~\ref{sec:coordination}}): 
consider providing notifications of collaborators' activity, since participants wanted to know of changes made by others in order to protect against destructive changes.
This is both a workaround for the limited field of view in current VR headsets, but also the ability to be aware of changes made behind one's back is a potential visualisation ``superpower'' \cite{willett2021perception}.

\noindent\textbf{Navigation} (\mrv{based on the discussion on \textit{Conversation} in Section~\ref{sec:activity}}): 
consider providing navigation mechanisms, to allow users to ``scroll'' the whole immersive space in all the three dimensions (\textit{x}, \textit{y}, \textit{z}). \mrv{Participants wanted the ability to move the surrounding space to access out-of-reach objects without modifying layout and to get an overview of the environment.} We can also consider providing a miniature space to give an overview and inform users about current field of view. 

\noindent\textbf{Environment} (\mrv{based on the discussion on \textit{Conversation} in Section~\ref{sec:activity}}): 
consider providing options for users to choose different types of environments, decorative elements and avatars, to support different collaborative scenarios and atmosphere, or to support management of territories (\mrv{as per the discussion on \textit{Territories} in Section~\ref{sec:observations}}). The ability to adapt working environment for various requirements of collaborative relationships and feelings can improve efficiency, effectiveness and engagement in collaboration~\cite{raziq2015impact}.  \mrv{For example, participants who spent time playing with avatars or gestures in VR, might have done so less in a more formal VR environment like a virtual classroom or conference room.}

\noindent\textbf{Automatic placement} (\mrv{based on the discussions on \textit{Interaction} in Section~\ref{sec:activity} and \textit{Spatial Organisation} in Section~\ref{sec:observations}}):
consider providing auto-alignment mechanisms to facilitate placing virtual objects, and providing automatic layout options to organise the objects. While layout and alignment tools are frequently found in desktop diagramming tools, in immersive environments they still reduce physical work required to place objects but are arguably more important because this work is increased due to the third dimension. For example, allowing snapping to 2D panels to facilitate alignment to planes and spherical surfaces, which were tasks we observed users doing manually in our study (Fig.~\ref{fig:teaser}).







%% file: 5.conclusion.tex
\section{Limitations, Conclusions and Future Work}
\label{sec:conclusion}
Our results suggest many positive outcomes and potential advantages for performing distributed collaborative sensemaking tasks that involve spatial organisation in VR over more standard Desktop environment.  Our main conjecture is that many of the advantages stem from the natural embodied interaction and communication that is possible in VR, and this would lead to increased engagement in the task. However, we note that there could also be other sources of increased engagement in VR, such as the isolation effect of VR headsets reducing environmental distractions, or the novelty effect of using VR compared to more familiar desktop communication and collaboration tools.

\mrv{
Our study employed basic and natural techniques that are suitable for novice users.
In the future, these techniques could be used as the basis for a suite of studies varying the parameters of the shared immersive environment, such as the introduction of table and wall surfaces, ranged interactions with objects, and navigation techniques such as teleport and zoom to enable larger shared virtual workspaces. 
Collaboration in immersive sensemaking is an area that has not been well studied.
This research involved groups of four subjects.
In the future, the measures that we have developed to investigate group user behaviours could be extended to fit larger group settings. Also, we used a variety of objective measures for behavioural analysis, but only one subjective measure, Social Presence. In the future, these measures could be supplemented with more subjective surveys around group dynamics (such as who the perceived leader was), and communication behaviours (such as subject perception of how easy it was to communicate with each other). It would also be good to use interviews to capture more overall feedback about the experience.
}

\mrv{
In summary, we believe that group-based immersive sensemaking is a rich area for future research, and hope that the measures we have developed could be of use to broader research community.
}

%% file: 6.appendix.tex
\section{Appendix}
\label{sec:appendix}

\subsection{Measure Definitions}
\label{app:definitions}

\subsubsection{Task}~\\
\indent \textbf{Accuracy.} The proportion of the number of images that were categorised into the correct emotion terms to the total number of images.

\textbf{Completion Time.} Between the time that the experimenter started showing the task content to the participants and the time that the experimenter changed the content to the next task.

\subsubsection{Interaction}~\\
\mrv{\indent \textbf{Grab.} The interaction that moves an object but does not change the grouping of an image (change the label to which an image is closest).}

\mrv{\textbf{Regroup.} The interaction that moves an object and changes the grouping of an image.}

\mrv{\textbf{Interaction Proportion.} The proportion of interaction duration in each task to the task completion time.}

\mrv{\textbf{Interaction Count.} The number of times that interactions are applied to objects.}

\textbf{Interaction Equality.} The balance between the duration of interaction that each participant contributed (equality of duration) and the balance between the interaction count of each participant (equality of count). 
The Gini coefficient~\cite{david1968miscellanea} that is a commonly used measure of inequality was used to do the calculation. As the sample size was not large, a refined bias-corrected method of the Gini coefficient~\cite{dixon1987bootstrapping} was used.

\subsubsection{Conversation}~\\
\indent \textbf{Conversation Proportion.} The proportion of conversation duration in each task to the task completion time. 

\textbf{Conversation Equality.} The balance between the duration of conversation that each participant contributed. Again the bias-corrected Gini coefficient was used to do the calculation.

\subsubsection{Coordination}~\\
\indent \textbf{Planning.} Planning refers to the high-level decisions about the tasks and collaborations. The discussions would relate to the strategies for organising the image and label cards, and dividing the responsibilities among participants.

\textbf{Assistance.} Participants help each other during collaboration. It is further divided into two categories: \textit{Question} (directly requesting help from others, asking/answering questions), and \textit{Exchange} (exchanging information). 

\textbf{Monitoring.} Gutwin and Greenberg~\cite{gutwin2000mechanics} defined monitoring as the awareness of others in the shared workspace, such as the information of \textit{who} are the others, \textit{where} are the others and \textit{what} the others are doing. \textit{Monitoring} here is considered as explicit observation of others and focus on others' activities during collaboration. The \textit{Monitoring} activity was retrieved automatically from audio and visual session playback when participants did not speak or interact with any objects longer than 5 seconds.

\textbf{Protection.} Protection refers to that participants defended their own work. This activity was retrieved automatically from visual playback and audio recordings, when participants made a response to the discussion of the images that they had grouped. 

\textbf{Respect.} Respect refers to a participant only modified other's work after asking for permission. This activity was captured automatically from visual playback and audio recordings. The criteria for retrieving this was when participants wanted to changed the grouping of an image, they asked others' opinions. 

\textbf{Accommodation.} Accommodation relates to dealing with conflicts during collaboration, such as resolving different opinions (e.g. \textit{Voting}) and interacting to the same object at the same time (e.g. \textit{Contention}). 

\subsubsection{Awareness}~\\
\indent \textbf{No Response.} No response when a participant initialised a question. 

\textbf{Think Aloud.} A participant talked to self.

\textbf{Conversation Conflict.} More than one conversation happened concurrently.

\textbf{Shaking.} When speaking, a participant shook the virtual hand (VR) / mouse cursor (desktop) or objects.

